\documentstyle[12pt]{article}

\renewcommand{\theequation}{\thesection.\arabic{equation}}

\def\res{\rm res}
\def\nkdvh{$n$--th KdV hierarchy}
\def\nmkdvh{$(n,m)$--th KdV hierarchy}

\def\bhs{bi--Hamiltonian structure}
\def\shs{second Hamiltonian structure}

\def\DS{Drinf'eld--Sokolov}
\def\pdo{pseudodifferential operator}
\def\pdos{pseudodifferential operators}
\def\baf{Baker--Akhiezer function}

\def\a{\begin{eqnarray}}
\def\b{\end{eqnarray}}
\def\0{\nonumber}
\def\ba{\begin{array}}
\def\ea{\end{array}}
\def\noal{\noalign{\vskip10pt}}

\def\ha{{1\over2}}

\def\lm{{\lambda}}
\def\dxy{\delta(x-y)}
\def\ddxy{\delta'(x-y)}
\def\tpsi{\widetilde\psi}

\def\ho{\hat {\bf {\cal H}}}

\def\l{{\cal L}}
\def\m{{\cal M}}

\def\LM{\Lambda}

\def\ro{{\cal R}}
\def\hp{{\hat {\cal P}}}
\def\hk{{\hat {\cal K}}}
\def\hm{{\hat {\cal M}}}
\def\thp{{\hat{\widetilde {\cal P}}}}
\def\tq{{\widetilde q}}
\def\tp{{\widetilde p}}

\def\htn{{\hat {\widetilde {\cal N}}}}

\def\lnm{{ L_{[n,m]} }}

\newlength{\extraspace}
\newlength{\extraspaces}
\newcounter{dummy}

\newcommand{\ai}{
\addtocounter{equation}{1}
\setcounter{dummy}{\value{equation}}
\setcounter{equation}{0}
\renewcommand{\theequation}{\thesection.\arabic{dummy}\alph{equation}}
\begin{eqnarray}
\addtolength{\abovedisplayskip}{\extraspaces}
\addtolength{\belowdisplayskip}{\extraspaces}
\addtolength{\abovedisplayshortskip}{\extraspace}
\addtolength{\belowdisplayshortskip}{\extraspace}}
\newcommand{\bj}{
\end{eqnarray}
\setcounter{equation}{\value{dummy}}
\renewcommand{\theequation}{\thesection.\arabic{equation}}}

\def\d{{\partial}}

\newcommand{\ddt}[1]{{\partial \over \partial t_{#1}}}

\newcommand{\bac}{\begin{array}{c}}
\newcommand{\bacc}{\begin{array}{cc}}
\newcommand{\baccc}{\begin{array}{ccc}}
\newcommand{\barcl}{\begin{array}{rcl}}
\newcommand{\bacccc}{\begin{array}{cccc}}
\newcommand{\baccccc}{\begin{array}{ccccc}}
\newcommand{\baccccccc}{\begin{array}{ccccccc}}
\newcommand{\barclcrcl}{\begin{array}{rclcrcl}}
\newcommand{\bacl}{\begin{array}{cl}}
\newcommand{\bal}{\begin{array}{l}}
\newcommand{\bacll}{\begin{array}{cll}}
\begin{document}
\begin{flushright}
BONN--TH--94--17\\
SISSA--ISAS--118/94/EP\\
AS--ITP--94--43\\
hepth/9408035
\end{flushright}

\centerline{\LARGE\bf The integrable hierarchy constructed from}
\vskip0.5cm
\centerline{\LARGE\bf a pair of KdV--type hierarchies }
\vskip0.5cm
\centerline{\LARGE\bf and its associated $W$ algebra}
\vskip1.5cm
\centerline{\large  L.Bonora}
\centerline{International School for Advanced Studies (SISSA/ISAS)}
\centerline{Via Beirut 2, 34014 Trieste, Italy}
\centerline{INFN, Sezione di Trieste.  }
\vskip0.5cm
\centerline{\large Q. P. Liu}
\centerline{Institute of Theoretical Physics, Chinese Academy of Science}
\centerline{P. O. Box 2735, 100080 Beijing, China}
\vskip0.5cm
\centerline{\large C.S.Xiong}
\centerline{Physikalisches Institut der Universit\"at Bonn}
\centerline{Nussallee 12, 53115 Bonn, Germany}
\vskip3cm
\abstract{For any two arbitrary positive integers `$n$' and `$m$',
using the $m$--th KdV hierarchy and the $(n+m)$--th KdV hierarchy as
building blocks,  we are able to construct another integrable hierarchy
(referred to as the $(n,m)$--th KdV hierarchy).
The $W$--algebra associated to the \shs\, of the $(n,m)$--th KdV hierarchy
(called $W(n,m)$ algebra) is isomorphic via a Miura map
to the direct sum of a $W_m$--algebra, a $W_{n+m}$--algebra and
an additional $U(1)$ current algebra. In turn, from the latter,
we can always construct a representation of a $W_\infty$--algebra.}

\vfill
\eject

\section{Introduction}

\setcounter{equation}{0}

Our purpose in this paper is to show how to construct new integrable
hierarchies starting from a couple of KdV--type hierarchies plus a $U(1)$
{\it gluon} current. Also in order to give the coordinates of our paper
with respect to the current literature, let us recall a few fundamental
things about KdV hierarchies.

There are two different description of the n--th  KdV hierarchy.
One is based on the so--called \pdo\, analysis (see \cite{dickey}),
in which we start from a differential operator $L$, called scalar Lax operator,
\a
L=\d^n+\sum_{i=1}^{n-1}u_i\d^{n-i-1}, \qquad \d=\frac{\d}{\d x}. \label{nkdvl}
\b
where the $u_i$'s are functions of the `space' coordinate $x$. Throughout the
paper the symbol $L$ will mean (\ref{nkdvl}).
After introducing the inverse $\d^{-1}$ of the derivative $\d$ (i.e. the formal
integration operator),
\a
&&\d \d^{-1}=\d^{-1} \d=1, \0\\
&& \d^{-1} f(x)= \sum_{l=0}^\infty (-1)^l f^{(l)}\d^{-l-1} \0
\b
we can calculate the fractional powers of $L$.

In general for a \pdo, $A=\sum_{i\leq n}a_i\d^i$, we define
\a
A_+=\sum_{i\geq0}a_i\d^i,\qquad A_-= A- A_+, \qquad \res (A)=a_{-1}.\0
\b
Since the operator $[(L^{r\over n})_+, L]=[L, (L^{r\over n})_-]$
is a purely differential operator of order $(n-2)$ for any positive integer
$r$, it naturally defines a series of infinite many differential equations
\a
\ddt r L=[(L^{r\over n})_+, L], \qquad r\geq 1,\label{nkdvh}
\b
where $t_1\equiv x$, while $t_2, t_3, \ldots,$ are real time parameters.
This set of equations is usually referred to as the \nkdvh or, simply, the
n--KdV hierarchy.

Another presentation of the n--KdV type hierarchy is by means of
the Drinf'eld--Sokolov
construction \cite{ds}. In such approach, we begin with a first order
matrix differential operator
\a
\l=\d+q-\LM, \label{mlgen}
\b
where $q$ and $\LM$ are $n\times n$ matrices and
\a
\LM=\lm E_{n1}+I,\quad I=\sum_{i=1}^{n-1}E_{i,i+1},\qquad
(E_{ij})_{kl}=\delta_{ik}\delta_{jl}\label{Lambda}
\b
where $\lm$ is the spectral parameter, while $q$ is a lower triangular matrix.
One can find a formal series
\a
T={\bf 1}+\sum_{i=1}^\infty T_i\LM^{-i},\0
\b
such that
\a
\l_{(0)}=T\l T^{-1}=\d+\LM +\sum_{i=0}^\infty f_i\LM^{-i},\0
\b
with all $f_i$ being functions. The centralizer
of $\l_{(0)}$ contains nothing but the constant elements of the
Heisenberg subalgebra, thus we can easily get the centralizer of the operator
$\l$, and we can define a series of flow equations
\a
\ddt r \l=[\m_r^+, \l], \qquad \m_r=T^{-1} \LM^r T. \label{ds}
\b
where the superscript ``$+$" means that we keep only non--negative
powers of $\LM$.
These equations are only  defined for classes of gauge equivalence, i.e.
up to transformations which leaves (\ref{ds}) form invariant.
If we suitably fix the gauge, eq.(\ref{ds}) reduces to the \nkdvh\,
(\ref{nkdvh}).

One can generalize the above construction in different directions.
In fact this
construction is based on the Lie algebra $sl_n$, and $\LM$ can be
understood as an element of the associated affine algebra which enjoys
particular properties.
The generalization in which
$\Lambda = \sum_{i=0}^{n-1}e_i$, where $e_i$ are the standard
Chevalley generators of an
affine Kac--Moody algebra (in this
case, $\Lambda$ is a grade one element of the principal Heisenberg subalgebra),
and $q$ is an element belonging to the relevant
non--positive graded Borel subalgebra, has been studied in \cite{ds},

Recently there have been several attempts to generalize the KdV--type
integrable hierarchies in other directions. One possibility is to replace
$\LM$ in (\ref{mlgen}) by any constant regular
element of any Heisenberg subalgebra of the Kac--Moody algebra.
The hierarchy constructed in such a way is called type I;
if the element $\LM$ is not regular the hierarchy  is called type II
\cite{ghm}.
It has been shown that, in the $gl_n$ case, the graded
regular elements exist only in some very special cases; furthermore, after
gauge fixing, this extended \DS\, hierarchy reduces to the Gelfand--Dickii
matrix hierarchy, which is a simple extension of (\ref{nkdvh})
obtained by replacing the
scalar Lax operator (\ref{nkdvl}) by a matrix valued one \cite{feher}.
On the other hand, there has been so far no detailed discussion
about type II integrable hierarchies, due to their complexity (see however
\cite{Pousa}).

Moving from a completely different starting point, in a recent paper
we have considered another type of extension: we have modified the
Lax operator (\ref{nkdvl}) by adding some suitable pseudodifferential terms;
in this way we have obtained a new integrable hierarchy, which we have
called the \nmkdvh,  \cite{bx1}. Actually these hierarchies are not an artifact
of ours. They naturally appear in two-- (and multi--) matrix models describing
2D gravity coupled to conformal matter. Two--matrix models are in fact
characterized by Toda lattice hierarchies.  There is
a duality between Toda lattice integrable hierarchies and differential
integrable hierarchies, \cite{bx2},\cite{bra}, which enables us to extract
KP--type differential hierarchies from the lattice hierarchy and vice versa.
Moreover the differential hierarchies obtained in this way can be reduced
to new hierarchies while preserving integrability. The full set of KP--type
integrable hierarchies obtained from the Toda lattice hierarchy together
with their integrable reductions turn out to fill up exactly the set of
the $(n,m)$--th KdV hierarchies.

In this paper we present the \nmkdvh {} from the point of view of extending
the KdV--type hierarchies:
given any two KdV type hierarchies, say
an $m$--th KdV hierarchy and an $(n+m)$--th KdV hierarchy, plus a $U(1)$
{\it gluon} current $J$, we show how to construct
another integrable hierarchy and that the latter is exactly the \nmkdvh.
Moreover
we will show that there exists a Miura map which establishes an isomorphism
between the $W$--algebra associated to the \shs\, of the
$(n,m)$--th KdV hierarchy to the direct sum of $W_{n+m}$--algebra
and $W_m$--algebra, as well as an additional $U(1)$ current algebra (when
$m=1$ the isomorphism is simply with the direct sum of a $W_{n+1}$--algebra
and the $U(1)$ current algebra.

The paper is organized as follows. In section 2, we review some well--known
facts about $W_n$ algebras. Our main results are presented in section 3,
where we first construct the $(n,m)$--th
KdV hierarchy from the $(n+m)$--th KdV hierarchy and $m$--th KdV hierarchy,
then we show that the $W(n,m)$--algebra is related to
$W_{n+m}\oplus W_m\oplus U(1)$ by a Miura map.
In section 4, we will analyze the conformal spectrum, and construct
the \DS\, representation of the \nmkdvh. We show that the $(n,m)$--th
KdV hierarchies  correspond in part to type I and in part to type II
generalized \DS\, hierarchy, \cite{ghm}.
Several examples and some remarks are presented in section 5.

\section{ The \nkdvh\, and $W_n$ algebra }

\setcounter{equation}{0}

In this section we will review some well--known results on
the \nkdvh\, and the $W_n$--algebra.
More precisely, we will show how to derive the $W_n$--algebra from :  1) the
n--th KdV hierarchy; 2) a suitable infinitesimal deformation of the
corresponding differential operator; 3) field--dependent gauge
transformations (or residual gauge symmetry).

\subsection{The \nkdvh}

There is a natural inner product on the \pdo\, algebra, defined by
\a
<A>\equiv\int\, dx ~\res(A).\label{innerpdo}
\b
which enables us to define two compatible Poisson structures
\a
\{f_X, f_Y\}_1(L)=<L[Y, X]>,\quad f_X(L)=<LX>,\quad X=
\sum_{i=1}^{n-1}\d^{-i}\chi_i,
\label{pb1}
\b
\a
\{f_X, f_Y\}_2(L)=<(XL)_+YL>-<(LX)_+LY>+\frac{1}{n}\int
 \d^{-1}\Bigl([L, X]_{(-1)}\Bigl)[L, Y]_{(-1)}.
\label{pb2}
\b
These two Poisson structures give rise to two Poisson bracket algebras
of the basic independent fields $u_i$
\a
&&\{u_i(x), u_j(y)\}_1=(\ho_1)_{ij}[u(x)]\dxy. \label{ho1}\\
&&\{u_i(x), u_j(y)\}_2=(\ho_2)_{ij}[u(x)]\dxy. \label{ho2}
\b
We call $\ho_1$ and $\ho_2$ Hamiltonian operators.
They are $(n-1)\times(n-1)$ matrix operators and only contain
the derivative $\d$ and the basic fields. In particular,
eq.(\ref{ho2}) is referred to as the $W_n$ algebra.

The conserved quantities (or Hamiltonians) have very simple form
\a
H_r=\frac{n}{r}<L^{r\over n}>, \qquad \forall r\geq1. \label{hr}
\b
They generate the Hamiltonian flows (\ref{nkdvh}) through the Poisson brackets.

\subsection{Infinitesimal deformations of the Lax operator}

The \nkdvh\, (\ref{nkdvh}) can be viewed
as the consistency conditions of the following  spectral evolution problem
\ai
&&L\psi=\lm\psi, \label{spec}\\
&&\ddt r\psi=(L^{r\over n})_+\psi.\label{evol}
\bj
The function $\psi(\lm, t)$ is usually referred to as \baf.

As observed in \cite{diz},
calculating a $W_n$--algebra is equivalent to finding two infinitesimal
differential operators $P$ and $Q$, such that
\a
\delta L =QL-LP \label{deltalpq}
\b
and $(L+\delta L)$ still has the same form as (\ref{nkdvl}).
This is equivalent to saying that
eq.(\ref{spec}) with vanishing spectral parameter $\lm=0$ is invariant
under the infinitesimal deformations
\a
\psi\longrightarrow \psi+\delta\psi,\qquad
L\longrightarrow L +\delta L \0
\b
with  $\delta L$ specified by eq.(\ref{deltalpq}), and
\a
\delta\psi = P\psi. \label{deltaPpsi}
\b
In other words, after such infinitesimal deformation, we still have
\a
(L+\delta L)(\psi+\delta\psi)=0. \label{scalinv}
\b

\bigskip
\noindent
{\it Proposition 2.1} : If we choose
\a
P = (YL)_+ - \frac{1}{n}Z, \qquad Q= (LY)_+ - \frac{1}{n}Z, \label{pq}
\b
where $Y=\sum_{i=1}^{n-1} \d^{i-n} \epsilon_i$ is an arbitrary infinitesimal
\pdo, and
\a
Z= \int^x \Bigl([L, Y]_{(-1)}\Bigl),\0
\b
then the deformation (\ref{deltalpq}) coincides with the one derived from
the second Poisson structure (\ref{pb2}), i.e.
\a
\delta u_i = \sum_{j=1}^{n-1} (\ho_2)_{ij}[u]\cdot\epsilon_{j}.
\label{deltaui}
\b

\bigskip

\noindent
{\it Proof } : Let $X$ and $Y$ be two arbitrary \pdos{} (with $Y$
infinitesimal), then the variation of the functional $f_X(L)$ under the
transformation generated by $f_Y(L)$ with respect to the second Poisson
structure is as follows
\a
\delta f_X(L) &=& \{f_X, f_Y\}_2(L) \0\\
&=& <X\Bigl( (LY)_+L-L(YL)_+\Bigl)>
-\frac{1}{n} <Z(LX-XL)>\0
\b
Let $X$ be independent of the basic fields $u_i$, we obtain
\a
\delta L = \Bigl((LY)_+-\frac{1}{n} Z\Bigl) L- L
\Bigl((YL)_+-\frac{1}{n} Z\Bigl) \0
\b
which is just the formula (\ref{deltalpq}) with the identification (\ref{pq}).
On the other hand,
eq.(\ref{deltaui}) is a direct consequence of (\ref{pb2}), so it
must be the solution of eq.(\ref{deltalpq}) -- remember that
\a
f_Y(L) = \sum_{i=1}^{n-1} \int u_i\epsilon_i,\qquad\delta f_X(L) =
\sum_{i=1}^{n-1} \int \delta u_i \chi_i\0
\b
This ends the proof.

Let us see a simple example. Choose $Y=\d^{1-n}\epsilon$. A straightforward
calculation shows that
\a
Z= -\frac{n(n-1)}{2}\epsilon',\qquad
P= \epsilon\d-\frac{n-1}{2} \epsilon', \qquad
Q=\epsilon\d+\frac{n+1}{2}\epsilon' \label{nconfpq}
\b
Plugging this into eq.(\ref{deltalpq}), and using eq.(\ref{deltaui}), we get
\a
\delta L = \sum_{i=1}^{n-1}\Bigl( (\ho_2)_{i1}[u]\cdot \epsilon\Bigl)\d^{n-i-1}
= (\epsilon\d +\frac{n+1}{2}\epsilon') L
 - L  (\epsilon\d - \frac{n-1}{2}\epsilon') . \label{deltalconf}
\b
In particular we find that $u_1$ satisfies the Virasoro algebra
\a
\{u_1(x), u_1(y)\}_2 = (c_n\d^3+ u_1(x)\d+ \d u_1(x))\dxy, \quad
c_n=\frac{1}{2}\left(\bac n+1 \\ 3\ea\right). \label{virn}
\b
Thus the deformation we considered is nothing but an infinitesimal
conformal transformation. Furthermore
\a
\delta\psi = ( \epsilon\d-\frac{n-1}{2} \epsilon')\psi, \qquad
\delta(L\psi) = (  \epsilon\d+\frac{n+1}{2} \epsilon' )(L\psi) \0
\b
indicates that $\psi$ and $L\psi$ are primary fields with
conformal weights $\frac{1-n}{2}$ and $\frac{n+1}{2}$, respectively.

We may write down the global form of this diffeomorphism
\a
x\longrightarrow f(x), \qquad L\longrightarrow
\Bigl(f'(x)\Bigl)^{-\frac{n+1}{2}}{\widetilde L}
\Bigl(f'(x)\Bigl)^{-\frac{n-1}{2}}
\label{diff}
\b
where ${\widetilde L}=\d^n+{\widetilde u}_1\d^{n-2}+\dots$. From this
transformation law, one can obtain the conformal properties of all
fields \cite{mg}.

\subsection{Drinf'eld--Sokolov representation}

The basic idea to construct a matrix version of the \nkdvh\, is to find a
vector space representation of the differential (or pseudodifferential)
operator algebra. For such a purpose, we may linearize the spectral
equation (\ref{spec}) by introducing $(n-1)$ supplementary
fields as follows
\a
\psi_1\equiv\psi, \qquad \psi_{i+1}=\d\psi_i=\d^i\psi, \qquad 1\leq i\leq n-1.
\label{psij}
\b
Define
\a
U=\sum_{i=1}^{n-1} u_i E_{n,n-i},\quad
\LM=\lm E_{n1}+I,\quad I=\sum_{i=1}^{n-1}E_{i,i+1}.\0
\b
Further denote by $\Psi$ and $\l$ the column vector
$(\psi, \psi_2, \ldots, \psi_{n})^t$ and
the $(n\times n)$ matrix operator $(\d+U-\LM)$, respectively.
Then the linear system (\ref{spec},\ref{evol}) can be rewritten in matrix form
as follows
\ai
&&\l\Psi=0; \label{lspec}\\
&&\ddt {ni+r} \Psi=\m_{ni+r}\Psi, \qquad i\geq0, \quad n-1\geq r\geq1
\label{levol}
\bj
where $\m_{ni+r}$ is a uniquely determined $n\times n$ matrix field,
which is a differential
polynomial in the $u_i$'s and only contains non--negative powers of $\lm$.
In particular, the first $n-1$ elements of its last column (these are the
important ones) take the form
\a
(\m_{ni+r})_{n-j,n-1} =\sum_{ 0\leq l \leq i} \lm^{i-l}
\frac{\delta H_{r+nl}}
{\delta u_j}, \qquad 1\leq j\leq n-1. \label{mrjn}
\b
It is perhaps worth giving a proof of eq.(\ref{levol}) and eq.(\ref{mrjn}).
To this end we first prove the following Lemma
\bigskip

\noindent
{\it Lemma 2.2 } : Let $(i,j,r)$  be non--negative integers, and let
$1\leq j,r\leq n-1$. Then
\a
\d^j (L^{{r\over n}+i})_+ =
\sum_{l=0}^{i+1} \Bigl((\d^j L^{{r\over n}+l-1})_-L\Bigl)_+ L^{i-l}
- \Bigl(\d^j (L^{{r\over n}+i})_-\Bigl)_+ + \Big(\d^j L^{{r\over n }-1}\Big)_+
L^{i+1} \label{decom}
\b

\bigskip

\noindent
{\it Proof } : Since the operator $\d^j (L^{{r\over n}+i})_+$ is a
purely differential operator, we have
\a
\d^j (L^{{r\over n}+i})_+ =
\d^j L^{{r\over n}} L^i
- \d^j (L^{{r\over n}+i})_-
= \Bigl((\d^j L^{{r\over n}} L^i\Bigl)_+
- \Bigl(\d^j (L^{{r\over n}+i})_-\Bigl)_+ \0
\b
The second term on the extreme right hand side has order less than $n$,
while the first term can be rewritten as
\a
 \Bigl((\d^j L^{{r\over n}} L^i\Bigl)_+
=  \Bigl(\d^j L^{{r\over n}-1}\Bigl)_+ L^{i+1}
+  \Bigl((\d^j L^{{r\over n}-1})_- L^{i+1}\Bigl)_+. \label{decom0}
\b
Now suppose that $l$ is also a positive integer and $l\leq i$. Due to
the fact that $L$ is a purely  differential operator, we obviously have
\a
\Bigl((\d^j L^{{r\over n}+l})_+L\Bigl)_- =0 \0
\b
and
\a
\Bigl((\d^j L^{{r\over n}+l})_-L\Bigl)_-
= \Bigl(\d^j L^{{r\over n}+l+1}\Bigl)_-. \0
\b
which immediately leads to the following recursion relation
\a
\Bigl((\d^j L^{{r\over n}+l})_-L^{i-l}\Bigl)_+
= \Bigl((\d^j L^{{r\over n}+l})_- L\Bigl)_+ L^{i-l-1}
+ \Bigl((\d^j L^{{r\over n}+l+1})_-L^{i-l-1}\Bigl)_+  \label{ident}
\b
Using this identity and eq.(\ref{decom0}), it is straightforward
to show that eq.(\ref{decom}) is true. This ends the proof of
Lemma 2.1.

\smallskip

In order to prove eq.(\ref{mrjn}), we write
\a
\Bigl((\d^j L^{{r\over n}+l-1})_-L\Bigl)_+
= \sum_{k=0}^{n-1} \alpha_{l,k}\d^k \0
\b
Then
\a
\alpha_{l,k} = \res\Bigl[
\Bigl((\d^j L^{{r\over n}+l-1})_-L\Bigl)_+ \d^{-k-1}\Bigl] \0
\b
In particular for $k=n-1$, we have
\a
&&\alpha_{l,n-1} = \res\Bigl[
\Bigl((\d^j L^{{r\over n}+l-1})_-L\Bigl)_+ \d^{-n}\Bigl] \0\\
&& = \res\Bigl[(\d^j L^{{r\over n}+l-1})_-L\d^{-n}\Bigl]
 = \Bigl(\d^j L^{{r\over n}+l-1}\Bigl)_{(-1)} \0
\b
On the other hand, from
\a
\delta H_{r+nl} = \int\, dx \res\Bigl( (\delta L)L^{{r\over n}+l-1}\Bigl)
 = \int\, dx \sum_{j=1}^{n-1} (\delta u_{n-j})
 \Bigl( \d^{j-1} L^{{r\over n}+l-1}\Bigl)_{(-1)} \0
\b
we get
\a
\frac{\delta H_{r+nl} } {\delta u_{n-j} }
= \Bigl( \d^{j-1} L^{{r\over n}+l-1}\Bigl)_{(-1)} = \alpha_{l,n-1}.
\b
Therefore
\a
&&\ddt {ni+r} \psi_j =
\ddt {ni+r} \d^{j-1}\psi = \d^{j-1} (L^{{r\over n}+i})_+ \psi \0\\
&& = \sum_{l=0}^{i} \lm^{i-l}
\Bigl((\d^{j-1} L^{{r\over n}+l-1})_-L\Bigl)_+ \psi
- \Bigl(\d^{j-1} (L^{{r\over n}+i})_-\Bigl)_+\psi
+\lambda^{i+1}\Big(\d^{j-1}L^{{r\over n}-1}\Big)_+\0\\
&& = \sum_{l=0}^{i} \lm^{i-l+1}
\frac{\delta H_{r+nl} } {\delta u_{n-j} } \psi_{n} +\ldots
\b
The dots contain auxiliary fields $\psi_k$ with $1\leq k\leq n-1$.
This ends the proof of eq.(\ref{mrjn}).

The consistency conditions of eqs.(\ref{lspec},\ref{levol}) give
rise to the \DS\, representation of the \nkdvh\,
\a
\ddt {r+ni} \l=[\m_{r+ni}, ~~~\l].\label{dsh}
\b
To be precise, the term \DS\, integrable hierarchy is utilized for
the flow equations (\ref{ds}) with reference to the operator ${\cal L}$
defined in (\ref{mlgen}). As noticed above, (\ref{ds}) are gauge dependent;
if we fix a suitable gauge we recover eq.(\ref{dsh}).

A specification is necessary at this point. The linearization of
eq.(\ref{spec}) is not unique.
For example, we may introduce supplementary fields in the following way
\a
&&\tpsi_1=\psi, \quad \tpsi_{i+1}=(\d+h_{i})\tpsi_i,
  \quad 1\leq i\leq n-1;\0\\
&&(\d+h_n)\tpsi_{n}=\lm\psi, \qquad -h_n=h_1+h_2+\ldots+h_{n-1}.\0
\b
Then we have another linearized spectral equation
\a
{\widetilde {\l}}{\widetilde \Psi}=0,\qquad {\widetilde {\l}}
=\d+{\widetilde U}-\LM,\qquad
{\widetilde U}=\sum_{i=1}^n h_iE_{ii}.\0
\b
The difference between these two linearizations is just a Miura map. In general
we call Miura transformation a (non--invertible) gauge transformation which
maps a
minimal set of independent coordinates onto another minimal set.
Therefore, modulo Miura transformations, the linearization is unique.
Hereafter we only focus our attention on the linearization
(\ref{lspec}).

{\it Example} : for the second flow, we have
\a
\m_2=(U-\LM)^2-U'
 +{2\over n}\sum_{i\geq j}\left(\bac i-1 \\ j-1\ea\right)u_1^{(i-j)}E_{ij}.\0
\b

The LHS of eq.(\ref{dsh}) is independent of $\lm$, therefore
we can set $\lm=0$ on both sides of eq.(\ref{dsh}). In the remaining part
of this section, we will show that in the case  $\lm=0$, the  spectral
equation completely determines the $W_n$--algebra.

\bigskip
\noindent
{\it Proposition 2.3} : $(i)$. The spectral equation
\a
\l_0 \Psi= 0, \qquad \l_0 = \l(\lm=0)= \d +U-I \label{lspec0}
\b
is invariant under the following infinitesimal transformations
\a
\Psi\longrightarrow G\Psi, \qquad \l_0 \longrightarrow G\l_0 G^{-1},\qquad
G=1+R. \label{gaug}
\b
where the infinitesimal matrix field $R$ satisfies
\a
\delta U=[R, ~~~ \d+U-I], \label{deltaj}
\b
and $(U+\delta U)$ has the same form as $U$.

\noindent $(ii)$. The elements of $R$
are polynomials of the basic fields $u_i$. In particular we can choose
\a
R_{n-j,n} = \frac {\delta F(L)} {\delta u_j} \equiv \epsilon_{j}, \qquad
1\leq j\leq n-1 \label{Rrjn}
\b
where $F$ is any infinitesimal linear functional.

Eq.(\ref{deltaj}) and the choice (\ref{Rrjn})
completely determines the structure of the $W_n$--algebra.

\bigskip

\noindent
{\it Proof } : The first statement follows directly from the
invariance (\ref{scalinv}), i.e. we have
\a
(\l_0+\delta \l_0)(\Psi+\delta\Psi)=0. \label{invl}
\b
This equation requires that $R$ satisfy eq.(\ref{deltaj}),
which in turn determines $R$ up to $(n-1)$ arbitrary elements.
In particular setting
\a
R_{n-j,n} = \sum_{i,r}\frac{\delta H_{r+ni}}{\delta u_j}\delta t_{ni+r} \0
\b
and comparing with (\ref{mrjn}), we get the $(r+ni)$--th Hamiltonian flow.

Let us come now to the second part of the proof.
We remark that the choice of the
independent elements of $R$ is not unique as long as the only requirement
is to recover a $W_n$--algebraic structure, no matter what the coordinates are.
For example, we can choose
the first row to be independent, then
the variation of \baf\, can be derived from eq.(\ref{gaug})
\a
\delta\psi= \sum_{j=1}^n R_{1j}\psi_{j-1} = P\psi, \qquad
P=\sum_{j=1}^n R_{1j}\d^{j-1}. \label{pqr}
\b
However choosing the first $n-1$ elements of the last column as independent,
as in (\ref{Rrjn}),
is of particular importance, because the relation (\ref{Rrjn}) leads to
coincidence with the second Poisson bracket (\ref{pb2}).

To see this point let $\delta\psi = P\psi$ be given by eq.(\ref{pq}).
We notice that
\a
&& \d^j P = \d^j \Bigl((YL)_+ -\frac{1}{n}Z\Bigl)
   = \d^j YL -\d^j (YL)_- -\frac{1}{n} \d^j Z \0\\
&& = \Bigl(\d^j YL -\d^j (YL)_- -\frac{1}{n} \d^j Z \Bigl)_+
   = (\d^j Y)_+ L + \Bigl((\d^j Y)_-L\Bigl)_+ -\Bigl(\d^j (YL)_-\Bigl)_+
 -\frac{1}{n} \d^j Z \0\\
&&= (\sum_{i=n-j}^{n-1} \d^{i+j-n} \epsilon_i) L +
 \sum_{i=1}^{n-j-1} (\d^{i+j-n}\epsilon_i L)_+
 -\Bigl(\d^j (YL)_-\Bigl)_+  -\frac{1}{n} \d^j Z, \quad 0\leq j\leq n-2\0
\b
which shows that
\a
\delta\psi_j = \d^{j-1}(\delta\psi) = \d^{j-1} P \psi
= (\sum_{i=n-j+1}^{n-1} \d^{i+j-n-1} \epsilon_i) \lm\psi +
\epsilon_{n-j}\psi_{n} +\ldots \0
\b
The first term disappears in the case $\lm=0$, the last term only contains
the auxiliary fields $\psi_{k}$ with $1\leq k\leq n-1$. Thus we obtain
\a
R_{j, n} = \epsilon_{n-j} = \frac{\delta f_Y(L)}{\delta u_{n-j} }\0
\b
which shows that the choice (\ref{Rrjn}) coincides with the choices made in
Proposition 2.1, i.e. with the second Poisson structure (\ref{ho2}), alias
the $W_n$ algebra. This ends the proof.

The definition of $W_n$--algebra contained in the above proposition
exhibits the intimate relation between $W$--algebra
and Kac--Moody current algebra\cite{feher2}.
Let ${\cal A}$ be a gauge field (WZNW current)
valued on some Lie algebra ${\cal G}$, and $G$ be an element of
the corresponding Lie group. Then the general gauge transformation
reads
\a
{\cal A} \longrightarrow G(\d + {\cal A} )G^{-1} \label{gkm}
\b
This gauge symmetry implies that the components of the ${\cal A} $ field
obey a Kac--Moody current algebra. After (partially) fixing the gauge
\footnote{ For $gl(n)$, the standard gauge fixing conditions
consist of restricting ${\cal A} $ to be of the form $(U-I)$ as exhibited
in eq.(\ref{lspec0}). This is equivalent to imposing a set of first
and second class constraints; after reduction \`a la Dirac
we obtain a $W_n$--algebra.},  the gauge symmetry (\ref{gkm})
will be reduced to the form (\ref{deltaj}). From this point of view we may call
the symmetry considered in Proposition 2.2 the {\it residual
gauge symmetry}. On the other hand, since the symmetry (\ref{gaug})
(characterized by the matrix field $R$) explicitly depends on the basic
fields $u_i$, we may also call it {\it field dependent gauge
transformation}. In practice, deriving $W$--algebra by solving
eq.(\ref{deltaj}) seems to be easier than calculating Poisson
bracket (\ref{pb2}) (for example, see \cite{poly}).

\bigskip
\noindent
\underline{\it Example} :
The functional $F=\int\, dx u_1(x)\epsilon(x)$ generates
the following field dependent gauge transformation
\a
R_{\rm conf}=\epsilon I -\epsilon' I_0+ \quad {\rm lower~triangular~part},
\label{rconf}
\b
where $I_0$ is a diagonal matrix with elements
$(I_0)_{ii}=\frac{n-2i+1}{2}$. One can easily check $R_{\rm conf}$
indeed leads to the diffeomorphism (\ref{nconfpq}).

\section{ The $(n,m)$-th KdV hierarchy and $W(n,m)$-algebra}

\setcounter{equation}{0}
\setcounter{subsection}{0}

In this section we will construct the $(n,m)$--th KdV hierarchy from
a pair of ordinary higher KdV hierarchies plus a $U(1)$ current $J$.
We also discuss the $W(n,m)$--algebra, which is the algebra
associated to the \shs\, of the $(n,m)$--th KdV hierarchy.

\subsection{ Constructing the $(n,m)$-th KdV hierarchy}

Our construction of the $(n,m)$--th KdV hierarchy is based on
the following Theorem.

\bigskip

\noindent
\underline{\it Theorem 3.1} : Let $A$ and $B$ be two purely
differential operators
\a
A = \d^{n+m} + \sum_{i=1}^{n+m-1} u_i \d^{n+m-i-1}, \qquad
B = \d^m - \sum_{i=1}^{m-1} v_i \d^{m-i-1}, \label{ab}
\b
and $J$ be a function of $x$. Define
\ai
&&\ddt 2 A = [ \Bigl(A^{\frac{2}{n+m} }\Bigl)_+, ~~A]
+ m\Bigl( (J\d+\frac{n+m+1}{2}J')A - A(J\d-\frac{n+m-1}{2}J')\Bigl);
  \label{ddt2u}\\
&&\ddt 2 B = [ \Bigl(B^{\frac{2}{m} }\Bigl)_+, ~~B]
+(n+m)\Bigl( (J\d+\frac{m+1}{2}J')B - B(J\d-\frac{m-1}{2}J')\Bigl);
  \label{ddt2v}\\
&&\ddt 2 J= \Bigl(\frac{4}{n(n+m)} u_1
 +\frac{4}{nm} v_1 + \frac{n+2m}{2}J^2\Bigl)'. \label{ddt2j}
\bj
This set of differential equations give rise to an integrable system.

\bigskip

\noindent
{\it Proof}. We will show that eqs.(\ref{ddt2u}--\ref{ddt2j}) admit
a Lax pair representation. First we observe that eqs.(\ref{ddt2u})
can be rewritten as
\a
\ddt 2 A &=& \Bigl[ (\d+ {m\over2}J)^2 +\frac{2}{n+m}u_1 - \frac{m^2}{4} J^2
+ \frac{m(n+m)}{2}J' \Bigl] A \0\\
 &-& A\Bigl[ (\d+ {m\over2}J)^2 + \frac{2}{n+m}u_1 - \frac{m^2}{4} J^2
- \frac{m(n+m)}{2}J' \Bigl]. \label{ddt2a[]}
\b
Similarly we re--express eqs.(\ref{ddt2v}) as
\a
\ddt 2 B &=& \Bigl[ (\d+ {{n+m}\over2}J)^2 -\frac{2}{m}v_1
 - \frac{(n+m)^2}{4} J^2 + \frac{(n+m)m}{2}J' \Bigl] B \0\\
 &-& B\Bigl[ (\d+ {{n+m}\over2}J)^2 - \frac{2}{m}v_1 - \frac{(n+m)^2}{4} J^2
- \frac{m(n+m)}{2}J' \Bigl]. \label{ddt2b[]}
\b
Next we introduce a pseudo--differential operator as follows
\a
\lnm=\phi^{-\frac{m}{2}}A\phi^{-{n\over2}}B^{-1}\phi^{\frac{n+m}{2}},
 \qquad (\ln\phi)'=-J. \label{lab}
\b
and expand $\lnm$ in the powers of $\d$
\a
\lnm=\d^n + \sum_{i=0}^\infty w_i\d^{n-i-1}, \label{wi}
\b
Due to the identity
\a
\phi^{-1}\d\phi =\d-J \0
\b
all the coefficients $w_i$ turn out to be differential polynomials of
the fields $u_i$, $v_j$ and $J$. For example
\a
&&w_0=0, \qquad w_1=u_1+v_1+{1\over4}nm(n+m)({1\over2}J^2+J'),\0\\
&&w_2=u_2+v_2+mJu_1+(n+m)Jv_1+nv_1' \0\\\noal
&& ~~~~~ +{1\over4}nm(n+m)\Bigl(\frac{n+2m}{6}J^3
+(n+m-1)JJ'+\frac{2n+m-3}{3}J^{''} \Bigl) \label{a12uvj}
\b
and so on. The RHS of eq.(\ref{ddt2j}) is a derivative. Therefore
we can extract the equation of motion of the field $\phi$
\a
\ddt 2 (\ln\phi)= -\Bigl(\frac{4}{n(n+m)} u_1
 +\frac{4}{nm} v_1 + \frac{n+2m}{2}J^2\Bigl). \label{ddt2phi}
\b
In this passage we have ignored possible integration constants: this is
part of the definition of $\phi$.

Now using eq.(\ref{ddt2a[]}) and eq.(\ref{ddt2b[]}),
as well as eq.(\ref{ddt2phi}), we can derive the equation of motion
of the operator $L_{[n,m]}$
\a
\ddt 2 \lnm =[ \d^2 + {2\over n} w_1, ~~~\lnm]. \0
\b
Noting that
\a
\d^2 + {2\over n} w_1 = (\lnm^{2\over n} )_+, \0
\b
we finally get the following Lax pair representation
\a
\ddt 2 \lnm=[ (\lnm^{2\over n})_+, ~~~\lnm]. \label{gddt2l}
\b
This ends the proof of integrability.

\smallskip

Two remarks are in order.

1) If we define a map (Miura map) as follows
\ai
&& L_1 = \phi^{-\frac{m}{2}} A \phi^{m\over2}, \label{l1}\\
&& L_2 = \phi^{-\frac{n+m}{2}} B \phi^{\frac{n+m}{2}} =
(\d-S_1)(\d-S_2)\ldots(\d-S_m), \label{l2}\\
&&S_1+S_2+\ldots+S_m= \frac{(n+m)m}{2}J
\bj
then
\a
\lnm=L_1L_2^{-1} = \d^n+\sum_{i=1}^{n-1}a_i\d^{n-i-1}+
  \sum_{l=1}^m a_{n+l-1}\frac{1}{\d-S_l}\ldots
  \frac{1}{\d-S_2}\frac{1}{\d-S_1}. \label{glax}
\b
This is exactly the Lax operator considered in \cite{bx1}.
The $a_i$ fields in the first sum coincide with $w_i$.
The full integrable hierarchy is
\a
\ddt r \lnm=[(\lnm^{r\over n})_+, ~~~\lnm]. \label{nmkdvh}
\b
Following \cite{bx1}, we will call it the $(n,m)$--th KdV hierarchy.
The $W$--algebra associated to its second Hamiltonian structure is
referred to as $W(n,m)$--algebra.

2) In our construction of the $(n,m)$--th KdV hierarchy,
the field $J$ plays an essential role. If we set $J=0$,
eqs.(\ref{ddt2u}) and eqs.(\ref{ddt2v}) are two
KdV--type of differential equations; the former is the second flow
equation of the $(n+m)$--th KdV hierarchy, the latter is the second flow of
the $m$--th KdV hierarchy. Therefore, the field $J$ behaves like
a {\it gluon}, which mediates the interaction between the $m$--th
KdV hierarchy and the $(n+m)$--th KdV hierarchy.

\subsection{ $W(n,m)$--algebra}

It has been shown that there exists a \bhs\, connected with the \pdo\,
(\ref{wi}) \cite{OS}. This implies that the $(n,m)$--th KdV hierarchy
possesses two compatible Poisson structures
\a
\{f_X, f_Y\}_1(\lnm)=<\lnm [X,Y]_\ro>,\label{gpb1}
\b
and
\a
\{f_X, f_Y\}_2(\lnm)&=&<(X\lnm)_+Y\lnm>-<(\lnm X)_+\lnm Y>\0\\
&+&{1\over n}\int [\lnm,Y]_{-1}\bigg(\d^{-1}[\lnm,X]_{-1}\bigg).
\label{gpb2}
\b
where
\a
[X, Y]_\ro\equiv \ha\Bigl([\ro X, Y]+[X, \ro Y]\Bigl)
=[X_+, Y_+]-[X_-, Y_-], \qquad \ro A\equiv A_+-A_-.\0
\b
These two Poisson structures lead to two infinite dimensional Poisson
algebras among the fields $w_i$, which are

1) a $W_{1+\infty}$--algebra
\a
\{w_i(x), w_j(y)\}_1= ({\hat {\cal W}}_1)_{ij}(x)\dxy, \label{gwinf1}
\b
and

2) a $W_\infty$--algebra
\a
\{w_i(x), w_j(y)\}_2= ({\hat {\cal W}}_2)_{ij}(x)\dxy. \label{gwinf2}
\b
Since in the $(n,m)$--th KdV hierarchy we only have $(n+2m-1)$
fundamental independent fields, these two infinite dimensional
Poisson algebras are realized  via two {\it finite} dimensional algebras
on the dynamical variables $(S_i; a_i)$, \cite{bx1}.
We denote the corresponding Poisson brackets by
\a
\{q_i(x), q_j(y)\}_1=(\ho_1)_{ij}(x)\dxy. \label{gho1}
\b
and
\a
\{q_i(x), q_j(y)\}_2=(\ho_2)_{ij}(x)\dxy. \label{gho2}
\b
$q_i$ are the components of a $(n+2m-1)$--dimensional vector
$q=(a_1,\dots,a_{n+m-1};S_1,\dots, S_m)$.
$\ho_1$ and $\ho_2$ are the appropriate Hamiltonian operators: they
are $(n+2m-1)\times(n+2m-1)$ matrix operators.
We are particularly interested in the second Poisson structure.
Let us rewrite the Hamiltonian operator $\ho_2$ in block form
\a
\left(\bacc \hp_1 & \hk_1 \\ \hk_2 & \hp_2 \ea\right),\0
\b
where $\hp_1$ and $\hp_2$ are $(n+m-1)\times(n+m-1)$ and $(m\times m)$
matrix operators. The antisymmetry of the Poisson bracket implies
\a
\hk_1^\dagger=-\hk_2\0
\b
where the superscript ``${}^\dagger$" means the following conjugation operation
\a
\d^\dagger=-\d,\qquad f^\dagger=f, \qquad (AB)^\dagger=B^\dagger A^\dagger,
\qquad (M^\dagger)_{ij}=M_{ji}^\dagger \0
\b
for any ordinary function $f$, differential operators $A,B$, and
matrix operator $M$.

\bigskip

\noindent
\underline{\it Theorem 3.2} : The Miura map
\a
\hm : q=(a_1,\ldots, a_{n+m-1}; S_1,\ldots, S_m) \longrightarrow
\tq=(u_1,\dots,u_{n+m-1};\,
v_1,\dots,v_{m-1}; \, J) \label{miura}
\b
transforms the second Hamiltonian structure into the following
block diagonal form
\a
\hm \ho_2(\hm)^\dagger=
\left(\baccc \thp_1 & 0 & 0 \\ 0 & \thp_2 & 0 \\ 0 & 0 & \thp_3
\ea\right) \label{blockho2}
\b
where $\thp_1$, $\thp_2$ and $\thp_3$ are the Hamiltonian operators of a
$W_{n+m}$, $W_m$--algebra and $U(1)$ current algebra, respectively.
In other words, the fields $u_i$ form a $W_{n+m}$ algebra, the $v_i$ form
a $W_m$ algebra and  $J$ a $U(1)$ algebra, respectively.
The remaining Poisson brackets vanish.
\bigskip

In order to prove this Theorem, we proceed first to prove the following

\bigskip

\noindent
{\it Proposition 3.3}:  Suppose that  a $W$--algebra be defined by
$(n+2m-1)$ basic fields $(u_1,\ldots, u_{n+m-1}$; $v_1,\ldots, v_{m-1};J)$,
which satisfy the following properties

\noindent
$(i)$. The fields $u_i(1\leq i\leq n+m-1)$ form a $W_{n+m}$--algebra
\def\hom{{\hat \Omega}}
\def\hsg{{\hat \sigma}}
\a
\{u_i, u_j\}_2=\hom_{ij}[u]\dxy.   \label{os1}
\b

\noindent
$(ii)$. The fields $(-v_j)(1\leq j\leq m-1)$ form a $W_m$--algebra
\a
\{v_i, v_j\}_2=\hsg_{ij}[-v]\dxy. \label{os2}
\b

\noindent
$(iii)$. The field  $J$ forms a $U(1)$ current algebra
\a
\{J, J\}_2=\frac{4}{nm(n+m)}\ddxy, \label{pbjj}
\b
while the the three groups of fields $u_i$, $v_i$ and $J$ mutually commute
\a
\{u_i, v_j\}_2 =0, \qquad \{u_i, J\}_2 =0, \qquad \{v_j, J\}_2 =0,\quad
\forall i,j. \0
\b
Then eqs.(\ref{ddt2u}-\ref{ddt2j}) are Hamiltonian equations ensuing from
the above Poisson brackets and the Hamiltonian $H= \int w_2(x) dx $,
where $w_2(x)$ is given by eqs.(\ref{a12uvj}).

\bigskip

\noindent
{\it Proof} : With respect to the $W$--algebra specified in the above
Proposition, the Hamiltonian $H$ generates the following equations
\ai
&&\ddt 2 u_i = \hom_{i2}[u]\cdot1 +m \hom_{i1}[u]\cdot J
  \label{ddt2ui}\\
&&\ddt 2 v_i = -\hsg_{i2}[-v]\cdot1 -(n+m) \hsg_{i1}[-v]\cdot J
  \label{ddt2vi}\\
&&\ddt 2 J= \Bigl(\frac{4}{n(n+m)} u_1
 +\frac{4}{nm} v_1 + \frac{n+2m}{2}J^2\Bigl)'. \label{ddt2jj}
\bj
Now we are going to show that eqs.(\ref{ddt2u}-\ref{ddt2j}) can be
re--expressed in this form.

We recall that the Lax pair form of the second flow
equations of the $m$--th KdV hierarchy is
\a
\ddt {b2} B = [ \Bigl( B^{\frac {2} {m}} \Bigl)_+, B].  \label{kdvb}
\b
with Lax operator $B$ given in eqs.(\ref{ab}). This flow is generated
by the second Hamiltonian $H_{b2}= -\int v_2(x) dx$ through the Poisson
structure (\ref{os2}), i.e.
\a
\ddt {b2} v_i =\{ v_i, H_{b2}\}_2 = -{\hat \sigma}_{i2}[-v]\cdot1, \0
\b
which leads to
\a
\ddt {2b} B = -\sum_{i=1}^{m-1} (\ddt {b2} v_i) \d^{m-i-1} =
\sum_{i=1}^{m-1} (\hsg_{i2}[-v]\cdot1) \d^{m-i-1}. \label{2ndb}
\b
Comparing eqs.(\ref{kdvb}) and (\ref{2ndb}), we get the following indentity
\a
\sum_{i=1}^{m-1} (\hsg_{i2}[-v]\cdot1) \d^{m-i-1}
 = [ (B^{\frac{2}{m}})_+,~~~ B]. \label{hsg2kdvb}
\b
On the other hand, in eq.(\ref{deltalconf}), if we choose
$\epsilon = J(x) {\widetilde \epsilon}$ with ${\widetilde \epsilon}$
being $x$--independent infinitesimal parameter, then we immediately get
another identity
\a
\sum_{i=1}^{m-1} \Bigl( \hsg_{i1}[-v] \cdot J\Bigl) \d^{m-i-1} =
(J\d+\frac{m+1}{2}J')B - B(J\d-\frac{m-1}{2}J'). \label{hsg1kdvb}
\b
These two identities, (\ref{hsg2kdvb}) and (\ref{hsg1kdvb}), tell us
that eq.(\ref{ddt2u}) is exactly the same as eq.(\ref{ddt2ui}).

Similarly for the $(n+m)$--th KdV hierarchy, we have
\a
\sum_{i=1}^{n+m-1} (\hom_{i2}[u]\cdot1) \d^{n+m-i-1}
 = [ (A^{\frac{2}{n+m} })_+,~~ A], \label{hom2kdva}
\b
and
\a
\sum_{i=1}^{n+m-1} \Bigl( \hom_{i1}[u] \cdot J\Bigl) \d^{n+m-i-1} =
(J\d+\frac{n+m+1}{2}J')A - A(J\d-\frac{n+m-1}{2}J'). \label{hom1kdva}
\b
which guarantees the coincidence between eq.(\ref{ddt2v}) and
eq.(\ref{ddt2vi}). This completes the proof of Proposition 3.3.

Proposition 3.3 means that
$W_{n+m}\oplus W_m \oplus U(1)$ is, modulo a Miura transformation, the \shs\,
of the \nmkdvh.  Due to the uniqueness of the second Poisson structure,
it is just the $W(n,m)$--algebra (\ref{gpb2}), up to a Miura transformation.
This completes the proof of Theorem 3.2.

As a direct consequence of Theorem 3.2, we have

\bigskip

\noindent
{\it Corollary 3.4}: Suppose that we have a $W$--algebra specified
in Proposition 3.3, introduce an infinite set of fields $w_i$'s
by eq.(\ref{wi}), then the $w_i$'s satisfies the infinite dimensional
Poisson algebras (\ref{gwinf1}) and (\ref{gwinf2}).

Let us make one final remark.
The relation between $\phi$ and $J$ is the typical vertex operator relation
that allows us to express interacting fields in terms of free fields in
chirally split 2D conformal field theories. In this
case $\phi$ plays the role of the vertex
operator and $J$ is the derivative of a free field.
Since it is well--known that $W_n$ algebras are representable by means of free
fields, this implies that $W(n,m)$ algebras can also be represented by means
of free fields.

\section{ \DS\, representation}

\setcounter{equation}{0}
\setcounter{subsection}{0}
In this section we will derive \DS\, representation of the
$(n,m)$--th KdV hierarchy, and analyze other ways to determine
the $W(n,m)$--algebra.

\subsection{Infinitesimal deformation of the Lax operator}

As we did in section 2, we first construct
the associated linear spectral system of eqs.(\ref{nmkdvh})
\a
&&\lnm\psi=\lm \psi,\label{gspec}\\
&&\ddt r\psi=(\lnm^{r\over n})_+\psi.\label{gevol}
\b
Once again $\psi$ is a \baf. We are going to show that calculating
$W(n,m)$--algebra is equivalent to finding two infinitesimal differential
operators $P$ and $Q$, such that
\a
\delta \lnm = Q\lnm-\lnm P \label{deltalpqg}
\b
and $(\lnm +\delta\lnm)$ has the same form as $\lnm$. This property
reflects the symmetry of the spectral equation (\ref{gspec})
at $\lm=0$, i.e.
\a
\delta\psi= P\psi,\qquad (\lnm+\delta\lnm)(\psi+\delta\psi)=0.
\b

\bigskip
\noindent
{\it Proposition 4.1} : Eq.(\ref{deltalpqg}) coincides
with the second Poisson structure (\ref{gpb2}), if we choose
\a
P = (Y\lnm)_+ - \frac{1}{n}Z, \qquad Q= (\lnm Y)_+ - \frac{1}{n}Z, \label{pqg}
\b
where $Y=\sum_{i=1}^\infty \d^{i-n} \epsilon_i$ is an arbitrary infinitesimal
\pdo, and
\a
Z= \int^x \Bigl([\lnm, Y]_{(-1)}\Bigl).
\b

\bigskip

The proof is the same as the proof of Proposition 2.1. Although
the formulas (\ref{deltalpqg}) and (\ref{pqg}) have the same form as
eq.(\ref{deltalpq}) and eq.(\ref{pq}), one should keep in mind
that the differential part of $Y$ now plays an important role.

In the case $Y=\d^{1-n}\epsilon$,  we have
\a
\delta \lnm= (\epsilon\d +\frac{n+1}{2}\epsilon') \lnm-
\lnm (\epsilon\d- \frac{n-1}{2} \epsilon').\0
\b
As a consequence, $a_1$ satisfies Virasoro algebra
\a
\{a_1, a_1\}=(c_n \d^3+ a_1\d + \d a_1)\dxy, \label{vira}
\b
moreover $\psi$ and $(\lnm\psi)$ transform like  primary fields with conformal
weights $(\frac{1-n}{2})$ and $(\frac{n+1}{2})$, respectively. In other words,
the scalar Lax operator (\ref{glax}) transforms covariantly under
diffeomorphisms
\a
\left\{\bal
x\longrightarrow f(x),\\\noal
\lnm\longrightarrow \Bigl(f'(x)\Bigl)^{-\frac{n+1}{2}}{\widetilde L}_{[n,m]}
    \Bigl(f'(x)\Bigl)^{-\frac{n-1}{2}}
\ea\right.
\b
where
\a
{\widetilde L}_{[n,m]}=\d^n+{\widetilde a}_1(x)\d^{n-2}+\ldots\0
\b
Obviously, under this transformation, the differential part and
pseudo--differential part do not interwine. In other words, they transform
separately in a covariant way. Therefore we have following results:

\noindent
$(i)$. The $a_l$ fields can be separated into two subsets;
the fields $a_l(1\leq l\leq n-1)$ (the first subset) have the conformal
properties as the fields of the ordinary $W_n$--algebra.

\noindent
$(ii)$. Each factor of the pseudodifferential part must be a conformal
covariant operator, which implies that $a_{n+l}(0\leq l\leq m-1)$
are primary fields, and
\a
\frac{1}{\d-S_j}\longrightarrow \Bigl(f'(x)\Bigl)^{\frac{n+2j-1}{2}}
\frac{1}{\d-{\widetilde S}_j}\Bigl(f'(x)\Bigl)^{-\frac{n+2j-3}{2}},\0
\b
which results in the following Poisson brackets
\a
\{a_1, S_j\}_2=(\frac{n+2j-1}{2}\d^2+S_j\d)\dxy.\label{a1sj}
\b
Summing over $j$, and recalling the definition of the field $J$, we get
\a
\{a_1, J\}_2=(\d^2+J\d)\dxy, \qquad \{a_1, \phi\}_2=\phi\ddxy.\label{a1phi}
\b
In other words, the field $\phi$ is a primary field with conformal
weight one. If we define
\a
w={1\over4}nm(n+m)({1\over2}J^2+J') \0
\b
then it is easy to see that
\a
\{w, a_1\}=(c_\phi \d^3+w\d +\d w)\dxy, \qquad
c_\phi= -{1\over4} nm(n+m). \label{vir3}
\b

\noindent
$(iii)$. The conformal properties of the operator $\lnm$ and the field $\phi$
imply that
\a
\left\{\bal
A\longrightarrow
\Bigl(f'(x)\Bigl)^{-\frac{n+m+1}{2}}{\widetilde A}
\Bigl(f'(x)\Bigl)^{-\frac{n+m-1}{2}},\\\noal
B\longrightarrow
\Bigl(f'(x)\Bigl)^{-\frac{m+1}{2}}{\widetilde B}
\Bigl(f'(x)\Bigl)^{-\frac{m-1}{2}}.
\ea\right.
\label{conab}
\b
We immediately recognize that the differential operators $A, B$
transform in the same way as the Lax operator of the ordinary KdV hierarchy
under diffeomorphisms. As a result, we have
\a
&&\{u_1, a_1\}=(c_{n+m}\d^3+u_1\d +\d u_1)\dxy, \qquad
 c_{n+m} = \ha\left(\bac n+m+1 \\ 3\ea\right), \label{vir1}\\
&&\{v_1, a_1\}=(-c_m\d^3 +v_1\d + \d v_1)\dxy, \qquad
c_m = \ha\left(\bac m+1 \\ 3\ea\right). \label{vir2}
\b
Combining (\ref{vir3}), (\ref{vir1}), and (\ref{vir2}), we get
(\ref{vira}), which guarantees consistency.
For $a_1 = u_1 + v_1 + w$ and the set of fields $(u_i)$, $(v_j)$ and $J$ are
mutually commutative. Eqs.(\ref{vira}), (\ref{vir3})
and (\ref{vir1},\ref{vir2}) are four copies of Virasoro algebras;
the total central charge
\a
c=c_{n+m} - c_m +c_\phi =c_n
\b
depends only on the order of Lax operator $\lnm$, no matter what $m$
is. The field $\phi$ just
tunes the conformal weight so that $\lnm$ is a conformal operator with
weight $n$.

\subsection{ \DS\, representation}

In order to construct the \DS\, representation of the $(n,m)$--th KdV
hierarchy, we introduce a set of auxiliary fields
\a
\psi_1=\psi,\quad \psi_{i+1}=\d\psi_i=\d^i\psi,\quad 1\leq i\leq n-1;
\qquad  (\d-S_{i+1})\psi_{-i}=\psi_{1-i},\quad 0\leq i\leq m-1.\label{psi}
\b
In this way, we obtain the linearized version of the spectral equation
(\ref{gspec})
\a
\l\Psi=0,\qquad \l=\d+U-\LM; \label{glspeclm}
\b
with
\a
U=-\sum_{i=0}^{m-1} S_{m-i}E_{-i,-i}
+\sum_{i=1}^{n+m-1}a_i E_{n, n-i},\label{gcan}
\b
and
\a
\LM=\lm E_{n,1}+\sum_{i=-m+1}^{n-1}E_{i,i+1}.\label{Lambda'}
\b
In order to derive the flow equation of $\Psi$,
we need the following Lemmas.

\bigskip
\noindent
\underline{\it Lemma 4.2} : Let $P$ be a differential operator $P=\sum_{i=0}^k
p_i\d^i$ with order $k$ smaller than $m$, then it is always possible to find
a set of $\tp_i$ which are differential polynomials of $p_i$ and $S_i$,
such that
\a
P=\tp_0 + \sum_{i=1}^k \tp_i(\d-S_{m-i+1})\dots(\d-S_{m-1})(\d-S_m).\0
\b

\bigskip

The proof is straightforward.

\bigskip
\noindent
{\it Lemma 4.3} : Let $(i,j,r)$ be non-negative integers and
$1\leq j,r\leq n-1$. Then
\a
\d^j (\lnm^{{r\over n}+i})_+ &=& (\d^j \lnm^{{r\over n}-1})_+ \lnm^{i+1}
-\Bigl( \d^j (\lnm^{{r\over n}+i})_- \Bigl)_+  \label{identg}\\
&+& \sum_{l=0}^i \Bigl[ \Bigl( (\d^j \lnm^{{r\over n}+l-1})_- \lnm \Bigl)_+
+ \Bigl( (\d^j \lnm^{{r\over n}+l-1})_+ \lnm \Bigl)_- \Bigl] \lnm^{i-l}.
\0
\b
Furthermore,
\a
\Bigl( \d^{j-1}\lnm^{{r\over n}+l-1}\Bigl)_{(-1)}  =
\frac{\delta H_{r+nl} } {\delta a_{n-j} }.
\b

\bigskip

The proof is similar to the proof of Lemma 2.2. The appearance of
the second term in the square bracket reflects the fact that $\lnm$
is not a purely differential operator. Since
\a
\ddt {r+ni} \psi_j = \d^{j-1}\ddt {r+ni} \psi = \d^{j-1}(\lnm^{r+ni})_+ \psi \0
\b
the above two Lemmas imply that the RHS can be represented linearly in
$\psi_k ~(-m+1\leq k\leq n)$, and the coefficient in front of  $\psi_{n}$
is $\sum_{l=0}^i \lm^{i-l} \frac{\delta H_{r+nl} } {\delta a_{n-j} }$.

Now let us turn our attention to the derivation of the equations of
motion for the elements $\psi_{-j}~(0\leq j\leq m-1)$.
{}From the definition (\ref{psi}), we have
\a
\ddt {r+ni} \psi_{-j}=
\Bigl(\sum_{l=1}^{j+1} \frac{1}{\d-S_j} \dots \frac{1}{\d-S_l}
\Bigl(\ddt {r+ni} S_l\Bigl) \frac{1}{\d-S_l} \dots \frac{1}{\d-S_1}
+\frac{1}{\d-S_{j+1}} \dots \frac{1}{\d-S_1} (\lnm^{{r\over n}+i})_+\Bigl)
\psi. \0
\b
Define $O_1=(\lnm^{r\over n})_+$, then eq.(\ref{nmkdvh}) guarantees
the following recursion relations
\a
(\d-S_i) O_{i+1} - O_i (\d- S_i) = \ddt {r+ni} S_i, \qquad 1\leq i\leq m \0
\b
or equivalently
\a
 O_{i+1} \frac{1}{\d-S_i} = \frac{1}{\d-S_i} O_i +
 \frac{1}{\d-S_i} \Bigl(\ddt {r+ni} S_i\Bigl) \frac{1}{\d-S_i}. \0
\b
These equations completely determine the operators $O_{i+1}$ and
the equations of motion $\ddt r S_i$. Applying this procedure, we are able
to get
\a
\ddt {r+ni} \psi_{-j} = O_{j+1} \frac{1}{\d-S_j} \dots \frac{1}{\d-S_1} \psi.
\0
\b
In the spirit of the above  Lemmas, we can rewrite the RHS of the
above expression
into the form $V_1 \cdot \Psi$, where the elements of row vector $V_1$
are Taylor series in $\lm$ and differential polynomials in $q$.
All these results together lead to the following Proposition

\bigskip

\noindent
\underline{\it Proposition 4.4} : $(i)$. Eq.(\ref{gevol})
can be uniquely rewritten in matrix form,
\a
\ddt {r+ni} \Psi=\m_{r+ni} \Psi \label{glevol}
\b
where the elements of $\m_{r+ni}$ are Taylor series in $\lm$, while they are
differential polynomials of the fields $(a_i, S_j)$, in particular
\a
(\m_{r+ni})_{n-j,n} =\sum_{ 0\leq l \leq i} \lm^{i-l} \frac{\delta H_{r+nl}}
{\delta a_j}. \label{gmrjn}
\b
$(ii)$ The consistency conditions of eqs.(\ref{glspeclm},\ref{glevol}) form
 the generalized \DS\, hierarchy
\a
\ddt {r+ni} \l=[\m_{r+ni}, ~~\l]. \label{gdsh}
\b

\bigskip
\noindent
\underline{\it Example} : The second flow corresponds to the following
matrix field
\a
\m_2=(U-\LM)^2+ {2\over n} \sum_{m+1\leq j\leq i\leq n+m}
\left(\bac i-m-1 \\ j-m-1\ea\right)a_1^{(i-j)}E_{ij}
-2\sum_{l=1}^{m-i+1}S_l' E_{ii}.\0
\b
Since $\LM$ in general is irregular, so the integrable hierarchy
(\ref{gdsh}) is type II generalized \DS\, integrable hierarchy
in the terminology of the reference 3.

\bigskip
\noindent
{\it Proposition 4.5} : $(i)$. The spectral equation with $\lm=0$
\a
(\d+U-I)\Psi=0. \label{glspec}
\b
is invariant under the following infinitesimal transformations
\a
\Psi\longrightarrow G\Psi, \qquad \l_0 \longrightarrow G\l_0 G^{-1},\qquad
G=1+R. \label{gaugg}
\b

\noindent
$(ii)$. The infinitesimal matrix field $R$ satisfies
\a
\delta U=[R, ~~~ \d+U-I], \label{gdeltaj}
\b
and $(U+\delta U)$ has the same form as $U$. The elements of $R$
are polynomials of the basic fields $u_i$'s, in particular
\a
R_{n-j,n} = \frac {\delta F} {\delta a_j} \equiv \epsilon_{j} \label{Rrjng}
\b
with $F$ being an arbitrary infinitesimal functional.

\noindent
$(iii)$. Eq.(\ref{gdeltaj}) completely determines the upper triangular part of
the matrix field $R$, and its main diagonal line (except the trace). Further,
\a
\delta a_i=\sum_{j=1}^m (\htn_1)_{ij} \delta S_j+
\sum_{j=1}^{n+m-1} (\htn_2)_{ij}\epsilon_j, \qquad 1\leq i\leq n+m-1.
\label{deltaas}
\b
where $\htn_1$ and $\htn_2$ are certain operatorial matrices of dimensions
$(n+m-1)\times m$, and $(n+m-1)\times(n+m-1)$ respectively.

\noindent
{\it Proof } : The first and the second statements can be
proved in the exactly the same way as  Proposition 2.3. In order to prove
the third statement,  we denote $R=\sum_{1-n-m}^{n+m-1} R_l$ and
$U=\sum_{1-n-m}^{n+m-1} U_l$, where $R_l(U_l)$ means the
pseudo--diagonal line $R_{i,j}(U_{i,j})$ with $i-j=l$. Then eq.(\ref{gdeltaj})
shows
\a
[R_l, I]=-R_{l+1}'+\sum_{k=1-n-m}[R_{l-k+1}, U_k], \qquad \forall l\geq0.\0
\b
Therefore we can recursively solve this equation for all $R_l(l\geq0)$,
which are differential polynomials in $(\epsilon_i, q_j)$. When $l< 0$,
eq.(\ref{gdeltaj}) means
\a
[R_l, I]=-\delta U_{l+1}-R_{l+1}'+\sum_{k=1-n-m}^0 [R_{l-k+1}, U_k],
\qquad \forall l<0, \label{rldeltaj}
\b
Solving these equations recursively, we obtain eq.(\ref{deltaas}).
This ends the proof.

Now let us suppose $F$ to be an arbitrary infinitesimal functional, define
\a
\frac{\delta F}{\delta a_j } = \epsilon_{j},
  \qquad (1\leq j\leq n+m-1), \qquad
\frac{\delta F}{\delta S_j}= \theta_j,\qquad 1\leq j\leq m.\0
\b
With respect to the second Poisson structure (\ref{gpb2}), $F$ generates
the infinitesimal transformation
\a
&&\delta a_i= \sum_{j=1}^{n+m-1} (\hp_1)_{ij}\epsilon_{j}
  +\sum_{j=1}^m (\hk_1)_{ij} \theta_j; \label{deltaai}\\
&&\delta S_i= \sum_{j=1}^{n+m-1} (\hk_2)_{ij}\epsilon_{j}
  +\sum_{j=1}^m (\hp_2)_{ij} \theta_j. \label{deltasi}
\b
Since the operatorial matrices $\hk_1, \hk_2$ are conjugate to each other,
there are only three independent matrix differential operators, say,
$\hp_1$, $\hp_2$ and $\hk_1$, which completely exhibit the structure of
the $W(n,m)$--algebra. However, as shown in the previous Proposition,
the field--dependent gauge symmetry can only determine the matrix operators
$\htn_1, \htn_2$ (in eq.(\ref{deltaas})).
Comparing eq.(\ref{deltaas}) and eqs.(\ref{deltaai},\ref{deltasi}), we
obtain
\a
\hp_1= \htn_1 \hk_2+ \htn_2, \qquad \hk_1 = \htn_1 \hp_2, \label{pkn}
\b
while $\hp_2$ remains undetermined. Therefore the field--dependent gauge
symmetry (\ref{gdeltaj}) is {\it necessary} but not sufficient
to completely determine the structure of the $W(n,m)$--algebra
(compare with subsection 2.2).
 This is quite a distinguished feature of the
$(n,m)$--th KdV hierarchy with $m\neq 0$.
We would like to point out that this is not a negative aspect.
In fact it is just this leftover arbitrariness which provides room
for the $W_\infty$--algebra.

\subsection{On the regularity properties of $\LM$}

Let us come now to the distinction mentioned in the introduction between
type I and type II integrable hierarchies according to whether
the constant element $\LM$ in the DS system is regular or not. We have
to study the regularity properties of the $\LM$'s defined in
eq.(\ref{Lambda'}).

We recall that and element $\Xi$, belonging to a
finite dimensional Lie algebra $\cal G$ or rank $r$, is regular if
\a
{\rm dim} ~{\cal G}(\Xi) = r \label{1}
\b
where
\a
{\cal G}(\Xi)= \{ X \in {\cal G}: (ad_\Xi)^k X =0 \quad\quad {\rm for~~some}~~k
= 1,2,\ldots\}\0
\b
It is known that, if $\Xi$ is regular, then ${\cal G}(\Xi)$
is a Cartan subalgebra
of ${\cal G}$.

Our elements $\LM$'s belong to the loop algebra (with loop parameter
$\lambda$):
the regularity property has to be understood with respect to their
projection on the relevant finite dimensional Lie algebra.

We have checked this regularity property for $\LM$, defined
by eq.(\ref{Lambda'}), by direct calculation
in the simplest cases, up to $n+m=5$ ($\LM$ is imbedded in $sl_{n+m}$
except when $n=1$, in which case it is understood to be imbedded in
$gl_{n+m}$).
It turns out that {\it $\LM$ is regular
if $m=0,1$, while it is not regular in the other cases}. We have further
studied the diagonalizability properties of $\LM$ in more general cases,
which confirms the above statement. This leads us to conjecture that
the above statement is true for any $n$ and $m$. We are therefore oriented
to believe that the $(n,m)$ hierarchies are type I if $m=0,1$, while
they are type II in the other cases.

Altogether we can conclude that the distinction between type I and II
integrable hierarchies has a technical meaning in the Drinf'eld--Sokolov
context, but does not seem to have any relevance whatsoever in the
Gelfand--Dickii context.

\section{Examples}

\setcounter{equation}{0}
\setcounter{subsection}{0}

In this section we present several examples to exhibit our construction
of the $(n,m)$--th KdV hierarchy, and to show the decomposition
of the $W(n,m)$--algebra into the direct sum.
For practical reasons we will proceed in reverse order with respect to
the demonstrations given so far. We will consider
the hierarchies and the $W(n,m)$--algebras explicitly
given in \cite{bx1}\cite{bx3}, and, subsequently, work out the
corresponding Miura
maps. We show that the modified Poisson algebras are a direct sum
of $W_m$, $W_{n+m}$ and $U(1)$, and the modified integrable
equations coincide with the ones given by eqs.(\ref{ddt2u}-
\ref{ddt2j}).

\noindent
\underline{ $(1,1)$-th KdV hierarchy}. The simplest case of the
$(n,m)$--th KdV hierarchy is with $n=m=1$. We choose it as our first
example because it has attracted a lot of attention both from mathematicians
\cite{math} and physicists \cite{tbr} in the past years.
The $(1,1)$--th KdV hierarchy is derived from one random matrix
model \cite{bx2}
\a
\ddt r L_{[1,1]} = [ (L_{[1,1]}^r)_+, ~~~L_{[1,1]}],\qquad
L_{[1,1]}= \d +a_1\frac{1}{\d-S_1}.
\b
The second flow equations are
\a
\ddt 2 a_1 =a_1^{''}+ 2(a_1S_1)', \qquad \ddt 2 S_1 = (2a_1 + S_1^2-S_1')'.
\b
The $W(1,1)$--algebra is
\a
\{a_1, a_1\}=(a_1\d + \d a_1)\dxy, \quad
\{a_1, S_1\}=(\d^2 +S_1\d)\dxy, \quad \{S_1, S_1\} =2\ddxy.
\label{w11}
\b
Define a map
\a
L_{[1,1]} = \phi^{-\ha} (\d^2 +u) \phi^{-\ha}  \d^{-1} \phi, \qquad
u= a_1 -\frac{1}{4}S_1^2 - \ha S_1', \qquad J= S_1, \label{11thmiura}
\b
then the field $u$ satisfies Virasoro algebra
\a
\{u, u\} = (\ha\d^3 +u\d +\d u)\dxy,\qquad \{J, J\} =2\ddxy.
\b
The flow equations become
\a
\ddt 2 u = \ha J^{'''} +2uJ' + u' J, \qquad
\ddt 2 J = 2u' +3JJ'. \label{11uvj}
\b
It is perhaps worth giving the third flow equations
\a
\ddt 3 u &=& u^{'''} + 6uu' +
  \frac{3}{4}(J^2)^{'''} +6uJJ' + \frac{3}{2}J^2u', \0\\
\ddt 3 J &=& (6Ju + \frac{5}{2} J^3 +J'')'.\0
\b
It is easy to see that this set of integrable equations extend
 the famous KdV equation by an additional
boson field. The Gelfand--Dickii Poisson brackets give
rise to two infinite dimensional Poisson algebras, which are referred to as
two boson representations of the $W_{1+\infty}$--algebra
and the $W_\infty$--algebra, respectively. This hierarchy has also a modified
Lax pair representation \cite{kuper}
\a
\ddt r L_{\rm mod} = [ \Bigl((L_{\rm mod}^r)_{\geq1}^\dagger\Bigl)^\dagger,
{}~~~L_{\rm mod} ]
\b
with the modified Lax operator
\a
L_{\rm mod} =\phi L_{[1,1]} \phi^{-1} = \d + J + h\d^{-1},\qquad
h=u+\frac{1}{4} J^2 +\ha J'.
\b
As a consequence, Gelfand--Dickii Poisson bracket should be modified
simultaneously\cite{das}.

\smallskip

\noindent
\underline{$(2,1)$-th KdV hierarchy}: This is the next simpler case.
The hierarchy is\cite{lx}\cite{bx3}
\a
\ddt r L_{[2,1]} = [(L_{[2,1]}^{r\over2})_+, ~~L_{[2,1]}],
\qquad
L_{[2,1]} = \d^2 + a_1 +a_2 \frac{1}{\d-S_1}\label{h21}
\b
The second flow equations are
\a
\ddt 2 a_1=2a_2'\qquad
\ddt 2 a_2=(a_2'+2a_2S_1)',\qquad
\ddt 2 S_1=(a_1+S_1^2-S_1')'. \label{f21}
\b
The Poisson algebra is
\a
&&\{a_1, a_1\}=(2a_1\d+a'_1+\frac{1}{2}\d^3)\delta(x-y),\qquad
\{a_1, a_2\}=(3a_2\d+2a'_2)\delta(x-y),\0\\
&&\{a_1, S_1\}=(\frac{3}{2}\d^2+S_1\d)\delta(x-y),\qquad
\{S_1, S_1\}=\frac{3}{2}\delta'(x-y), \label{w21}\\
&&\{a_2, a_2\}=[(2a'_2+4a_2S_1)\d
+a^{''}_2+2(a_2S_1)']\delta(x-y),\0\\
&&\{a_2, S_1\}=\Bigl(a_1\d+(\d+S_1)^2\d\Bigl)
\delta(x-y).\0
\b
The relevant map is
\a
L_{[2,1]} = \phi^{-\ha} (\d^3 +u_1\d +u_2) \phi^{-1}  \d^{-1} \phi^{3\over2}\0
\b
with
\a
J = \frac{2}{3} S_1, \quad
u_1 = a_1 - {1\over3} S_1^2 - S_1', \quad
u_2 = a_2 - {2\over3} a_1 S_1 - {2\over27}S_1^3 - {2\over3} S_1S_1'
- {2\over3}  S_1^{''}. \label{21thmiura}
\b
Then the fields $(u_1,u_2)$ satisfy the $W_3$--algebra
\a
&&\{u_1, u_1\} = (2\d^3+ u_1\d + \d u_1)\delta(x-y),\0\\
&&\{u_1, u_2\} =\Big(u_2\d+ 2\d u_2-\d^2 u_1-\d^4\Big)\delta(x-y),\label{w3}\\
&&\{u_2, u_2\} =[ \d^2 u_2 -u_2\d^2 -\frac{2}{3}(u_1+\d^2)(\d u_1+\d^3)]
\delta(x-y).\0
\b
Eq.(\ref{f21}) becomes
\a
&&\ddt 2 u_1=2u_2'-u_1'' + 2J^{'''} + 2u_1 J' + u'_1J \0\\
&&\ddt 2 u_2=u_2^{''}-\frac{2}{3}(u_1u_1'+u_1''') + J^{''''}
+u_1 J^{''} +3u_2J'+u_2'J \label{boussdilaton}\\
&&\ddt 2 J = \frac{2}{3}u_1' + 4JJ'. \0
\b
Once again this coincides with eqs.(\ref{ddt2u}-\ref{ddt2j}).
This equation is an extended version of the Boussinesq equation
(in parametric form) by the addition of one further boson
field. The various free field representations of the Poisson algebra have
been given in \cite{lx}.

\smallskip

\noindent
\underline{$(1,2)$-th KdV hierarchy} : This hierarchy has been studied
in \cite{bx3}.
\a
\ddt r L_{[1,2]} = [(L_{[1,2]}^r)_+, ~~L_{[1,2]}],
\qquad
L_{[1,2]} = \d + a_1{1\over{\d-S_1}}+a_2{1\over{\d-S_2}}{1\over{\d-S_1}}.
\label{h12}
\b
The second flow equations are
\a
&&\ddt 2 a_1=a_1^{''}+2a_2'+2(a_1 S_1)'\0\\
&&\ddt 2 a_2=a_2^{''}+2a'_2S_2+2a_2(S_1+S_2)'\label{3.1}\\
&&\ddt 2 S_1=2a'_1+2S_1S_1'-S_1^{''}\0\\
&&\ddt 2 S_2=2a'_1+2 S_2 S_2'-S_2^{''}-2 S_1^{''}\0
\b
The $W(1,2)$--algebra is
\a
&&\{a_1, a_1\}_2=(2a_1\d+a'_1)\delta(x-y),
\quad\quad\{a_1, a_2\}_2=(3a_2\d+2a'_2)\delta(x-y),\0\\
&&\{a_1, S_1\}_2=(\d^2+S_1\d)\delta(x-y),
\quad\quad\{a_1, S_2\}_2=(2\d^2+S_2\d)\delta(x-y),\0\\
&&\{a_2, a_2\}_2=[(2a'_2+4a_2S_2-2a_2S_1)\d
+a^{''}_2+(2a_2S_2-a_2S_1)']
\delta(x-y),\label{PB2}\\
&&\{a_2, S_2\}_2=\Bigl(a_1\d+(\d+S_2)(\d+S_2-S_1)\d\Bigl)
\delta(x-y),\0\\
&&\{S_1, S_1\}_2=2\delta'(x-y),\quad\quad\quad\{a_2, S_1\}_2=0,\0\\
&&\{S_1, S_2\}_2=\delta'(x-y), \quad\quad\quad\{S_2, S_2\}_2
=2\delta'(x-y),\0
\b
The Miura map we need is
\a
L_{[1,2]} = \phi^{-1} (\d^3 +u_1\d +u_2) \phi^{-\ha}(\d^2 -v_1)^{-1}
 \phi^{3\over2}\0
\b
with
\a
&&J =\frac{1}{3}(S_1+S_2),\qquad
v_1 = {1\over4} (S_2-S_1)^2+\ha (S_2-S_1)' \0\\
&&u_1 = a_1 - \frac{1}{3}(S_1^2+S_2^2-S_1S_2) -S_2' \\
&& u_2 = a_2 + \frac{1}{3}a_1(S_1-2S_2)
+ \frac{1}{27}(S_1+S_2)(5S_1S_2-2S_1^2-2S_2^2) \0\\
&&~~~-\frac{1}{3}(S_1-2S_2) (S_1-S_2)'
+\frac{1}{3}(S_1-2S_2)''.\0
\b
We remark that this is the first non--trivial Miura map we encountered
till now, since the maps (\ref{11thmiura}) and (\ref{21thmiura}) are
invertible, so as to be just redefinitions of the fields.
The fields $(u_1,u_2)$ satisfy the $W_3$--algebra (\ref{w3}), while
$v_1$ satisfies the Virasoro algebra
\a
\{v_1, v_1\} = (-\ha\d^3+ v_1\d + \d v_1)\delta(x-y)
\b
with negative central charge. The Poisson bracket of the field $J$ is
\a
\{J,~ J\} = \frac{2}{3}\ddxy.
\b
The modified second flow equations are
\a
&&\ddt 2 u_1=2u_2'-u_1'' + 4J^{'''} + 4u_1 J' +2u'_1J \0\\
&&\ddt 2 u_2=u_2^{''}-\frac{2}{3}(u_1u_1'+u_1''') +2J^{''''}
+2u_1 J^{''} +6u_2J'+2u_2'J \0\\
&&\ddt 2 v_1 =  -\frac{3}{2}J^{'''} + 6v_1 J' +3v_1' J \label{Bouss}\\
&&\ddt 2 J = \frac{4}{3}u_1' +2v_1' + 5JJ'. \0
\b
This  set of equations describe the coupling of Boussinesq equation
and KdV equation. It is worthwhile studying further.

In this paper, we have shown how to construct a new integrable
hierarchy from two KdV hierarchies, in particular we shown that
the corresponding $W(n,m)$--algebra can be decomposed into
a direct sum of the ordinary $W_n$--algebras. The Lie algebra structure
of the $W(n,m)$--algebra has been discussed in \cite{yu}. There are still
many problems which should be understood better. At first
it has been noticed
that the $W(1,1)$--algebra can be constructed from $sl(2)$ Kac-Moody
algebra through coset construction\cite{toppan}. It is interesting
to check if this is true for all the other $W(n,m)$--algebras.
Second, Dickey has observed that the $(n,1)$--th KdV can be viewed as
a reduction of KP hierarchy by fixing the additional symmetry \cite{dic}.
It is perhaps also true for $m>1$ case.
Finally the $(n,m)$--th KdV hierarchy might play roles in
the study of the low--dimesional quantum gravity.

\vskip1cm
{\bf Acknowledgements.} One of us (L.B.) would like to thank L.Dabrowski and
C.Reina for helpful discussions.

\end{document}